\documentclass[superscriptaddress,aps,pra,twocolumn,showpacs,nofootinbib,longbibliography,notitlepage,10pt]{revtex4-2}
\usepackage{etex}
\usepackage{Physics}
\usepackage{amsmath,amssymb,amsthm,mathrsfs}
\usepackage[colorlinks=true,citecolor=blue,urlcolor=blue]{hyperref}
\usepackage{subfigure}
\usepackage[pdftex]{graphicx}
\usepackage{times,txfonts}
\usepackage{braket}
\usepackage{color}
\usepackage{natbib}
\usepackage{amsmath,blkarray}
\usepackage{mathtools}
\usepackage{latexsym}
\usepackage{tabularx, booktabs}
\usepackage{graphics,epstopdf}
\usepackage{graphicx}
\usepackage{float}
\usepackage{graphicx}
\usepackage{amsfonts}
\usepackage{color,soul}

\begin{document}
	
	\title{Revealing Incommensurability between Device-Independent Randomness, Nonlocality, and Entanglement using Hardy and Hardy-type Relations}
	
	\author{Souradeep Sasmal}
	\affiliation{Raman Research Institute, Sadashivanagar, Bangalore, India}
	\affiliation{Indian Institute of Technology Hyderabad, Kandi, Sangareddy, Telengana, India}
	
	\author{Ashutosh Rai}
	\affiliation{School of Electrical Engineering, Korea Advanced Institute of Science and Technology (KAIST), 291 Daehak-ro, Yuseong-gu, Daejeon 34141, Republic of Korea}
	
	\author{Sayan Gangopadhyay}
	\affiliation{Department of Physics, Indian Institute of Science, C.V. Raman Avenue, Bangalore 560012, India}
	\affiliation{Institute for Quantum Computing, Department of Physics and Astronomy, University of Waterloo, Canada N2L3G1}
	
	\author{Dipankar Home}
	\affiliation{Centre for Astroparticle Physics and Space Science (CAPSS), Bose Institute, Block EN, Sector V, Salt Lake, Kolkata 700 091, India}
	
	\author{Urbasi Sinha}
	\email{usinha@rri.res.in}
	\affiliation{Raman Research Institute, Sadashivanagar, Bangalore, India}

	\begin{abstract}
		
	A comprehensive treatment of the quantification of randomness certified device-independently by using the Hardy and Cabello-Liang-Li (CLL) nonlocality relations is provided in the two parties – two measurements per party – two outcomes per measurement (2-2-2) scenario. For the Hardy nonlocality, it is revealed that for a given amount of nonlocality signified by a particular non-zero value of the Hardy parameter, the amount of Hardy-certifiable randomness is not unique, unlike the way the amount of certifiable randomness is related to the CHSH nonlocality. This is because any specified non-maximal value of Hardy nonlocality parameter characterises a set of quantum extremal distributions. Then this leads to a range of certifiable amounts of randomness corresponding to a given Hardy parameter. On the other hand, for a given amount of CLL-nonlocality, the certifiable randomness is unique, similar to that for the CHSH nonlocality. Furthermore, the tightness of our analytical treatment evaluating the respective guaranteed bounds for the Hardy and CLL relations is demonstrated by their exact agreement with the Semi-Definite-Programming based computed bounds. Interestingly, the analytically evaluated maximum achievable bounds of both Hardy and CLL-certified randomness have been found to be realisable for non-maximal values of the Hardy and CLL nonlocality parameters. In particular, we have shown that even close to the maximum 2 bits of CLL-certified randomness can be realised from non-maximally entangled pure two-qubit states corresponding to small values of the CLL nonlocal parameter. This, therefore, clearly illustrates the quantitative incommensurability between randomness, nonlocality and entanglement.
		
	\end{abstract}
	
	\maketitle
	
	\section{Introduction}
	
	Certification and quantification of reliable randomness as a resource for myriad applications in diverse areas is a cutting-edge topic of much interest. In this context, a remarkable realisation has been that violation of the CHSH inequality \cite{chsh} for the entangled states, apart from signifying nonlocality, also provides statistically verifiable device-independent (DI) certification of randomness, i.e., randomness is then guaranteed even for an imperfect or a tampered random number generating device \cite{Masanes2006, Pironio2010, Colbeck2012}. Nonlocality and DI certified randomness emerging as a concomitant feature of the CHSH inequality is intriguing and has inspired probing deeper into the nature of the relationship between them. In particular, the question arises as to whether the aforesaid DI certified randomness, nonlocality and entanglement are quantitatively commensurate in the sense that greater/smaller amounts of nonlocality and entanglement necessarily imply larger/smaller amounts of randomness. In this regard, it has been shown that the CHSH-certified guaranteed bound of randomness is monotonically related to nonlocality \cite{Pironio2010}. In contrast, using a tailor-made tilted-Bell inequality, it has been shown \cite{Acin2012} that close to the theoretical maximum of 2 bits amount of randomness can be certified from a maximally entangled two-qubit state having the CHSH violation tending to zero. This result of achieving close to 2 bits of certifiable randomness from maximally entangled two-qubit state has recently been made \cite{Colbeck2022} more robust by using different forms of tilted-Bell inequality having a wide range of CHSH values. Such results bring out the incommensurability between randomness and nonlocality in the two parties – two measurements per party – two outcomes per measurement (2-2-2) scenario. On the other hand, the incommensurability between randomness and entanglement has also been shown \cite{Acin2012} by achieving maximum amount of certifiable randomness from a pure non-maximally entangled state, but by going beyond the 2-2-2 scenario. Thus, beyond the 2-2-2 scenario, a line of study has been developed for demonstrating maximum amount of certifiable randomness by using different methods, such as increasing the number of measurement settings \cite{Andersson2018, Borkala2022, Mahato2022}, introducing higher outcome POVM \cite{Woodhead2020}.
	
	Against the above backdrop, the question that remains yet uninvestigated is whether it is possible to achieve close to 2 bits of certifiable randomness from a pure non-maximally entangled two-qubit state in the 2-2-2 scenario, which would enable showing in this simplest context, the incommensurability between randomness, nonlocality and entanglement in a single setup. To this end, in the present work, we have invoked different forms of local realist inequality other than the CHSH or tilted-CHSH inequality, introduced by Hardy \cite{Hardy1992} and Cabello, Liang and Li (CLL)~\cite{Cabello2002, Liang2005}, as means for generating DI certified randomness. In particular, we have come up with a strategy for achieving close to 2 bits of certified randomness from pure non-maximally entangled states in the 2-2-2 scenario using the CLL relations, thereby evidencing the incompatibility between randomness, nonlocality and entanglement in a single setup. Moreover, in this process of quantifying the certified randomness, we have found that, unlike the CHSH or tilted-Bell inequality cases, a given amount of Hardy-nonlocality corresponds to a set of quantum extremal distributions, and thus, the amount of certified randomness corresponding to a given amount of nonlocality is not unique. In other words, there exists a range of Hardy-certified randomness for a given value of the Hardy nonlocality parameter.
	
	To set the stage for our treatment, we begin (Sec.~\ref{certi}) by outlining the logical basis for regarding the validity of the Hardy and CLL relations as certifying DI certified randomness. For this purpose, the incompatibility of Hardy and CLL relations with the statistical condition of \textit{`predictability'} is shown by invoking the fundamental physical principle of \textit{`no signalling'} at the statistical (operational) level. As a consequence, the violation of in-principle `predictability' implying DI certified randomness is guaranteed by the empirical validity of the Hardy/CLL relations. A similar demonstration for the CHSH inequality was provided earlier \cite{Cavalcanti2012}. It is worth stressing that the certification of DI certified randomness in this way is independent of quantum theory as well as of who uses randomness. In contrast, the estimation of the amount of certified randomness depends on the theory as well as on the information available about the random number generator used and its trustworthiness.
	
	Next, in Sec.~\ref{qcgr}, by suitably quantifying the amounts of Hardy/CLL-certified randomness in terms of the guaranteed and the maximum achievable bounds, both the bounds are analytically evaluated for both quantum theory and no-signalling theory (Sec.~\ref{digr}). Then, we compare the analytically obtained quantum bound with the bound that has been numerically evaluated by employing the technique of Semi Definite Programming (SDP) \cite{Pironio2010, SDP1, SDP2}. The implications of these results and future directions of studies are discussed in the final Sec. \ref{con}. 
	

	\section{Certification of DI certified randomness using Hardy and Cabello-Liang-Li relations} \label{certi}
	
	First, we recall that derivation of the CHSH inequality from the assumption of predictability and the fundamental physical principle of no-signalling at the operational/statistical level \cite{Cavalcanti2012} provides a compelling justification for regarding the violation of CHSH inequality as falsifying predictability, thereby certifying DI certified randomness. We will now indicate the way similar arguments also hold good for the Hardy and CLL relations. 
	
	To put it precisely, in the context of the EPR-Bohm setup involving two space-like separated parties, say, Alice and Bob, the assumptions of predictability and the no-signalling condition used here at the operational level are as follows:
	
	(a)~\textit{Predictability:} Given any state preparation procedure $\kappa$, if the outcomes $a$ and $b$ of the measurements $\mathcal{A}_{x}$ and $\mathcal{B}_{y}$ of Alice and Bob respectively are predictable with \textit{certainty}. This means that the predicted probability of joint measurement outcomes is given by
	\begin{equation} \label{p}
		P(a, b|\mathcal{A}_{x},\mathcal{B}_{y},\kappa) \in \{0,1\} \ \ \ \forall \ \ a, b, \mathcal{A}_{x}, \mathcal{B}_{y}, \kappa
	\end{equation}
	
	(b)~\textit{No-signalling condition:} The observable probability of the occurrence of any measurement outcome in any one of the two wings of the setup is independent of the choice of the measurement setting in the other wing, i.e.,
	\begin{eqnarray}
		P(a|\mathcal{A}_{x},\mathcal{B}_{y},\kappa) &=& P(a|\mathcal{A}_{x},\kappa) \ \ \ \forall \ \ a, \mathcal{A}_{x}, \mathcal{B}_{y}, \kappa \label{ns1} \\
		P(b|\mathcal{A}_{x},\mathcal{B}_{y},\kappa) &=& P(b|\mathcal{B}_{y},\kappa) \ \ \ \forall \ \ b, \mathcal{A}_{x},\mathcal{B}_{y}, \kappa \label{ns2}
	\end{eqnarray}
	
	The first step in the argument is that the above stated conditions embodied in Eqs. (\ref{p}-\ref{ns2}) lead to the following condition of factorisability of the joint probabilities of measurement outcomes at the operational/statistical level \cite{Cavalcanti2012}
	\begin{equation}\label{fact}
		P(a, b|\mathcal{A}_{x},\mathcal{B}_{y},\kappa) = P(a|\mathcal{A}_{x},\kappa) \ P(b|\mathcal{B}_{y},\kappa) \ \ \ \forall \ \ a, b, \mathcal{A}_{x}, \mathcal{B}_{y}, \kappa
	\end{equation}
	
	Then the key point is that the simultaneous validity of the Hardy relations in the 2-2-2 scenario given by
	\begin{eqnarray} 
		P(+1,+1|\mathcal{A}_{1},\mathcal{B}_{1}, \kappa) &=& \mathcal{P}_{Hardy} > 0 \label{h1}\\[2.5pt]
		P(-1,+1|\mathcal{A}_{2},\mathcal{B}_{1}, \kappa) &=& 0 \label{h2}\\[2.5pt]
		P(+1,-1|\mathcal{A}_{1},\mathcal{B}_{2}, \kappa) &=& 0 \label{h3}\\[2.5pt]
		P(+1,+1|\mathcal{A}_{2},\mathcal{B}_{2}, \kappa) &=& 0 \label{h4}
	\end{eqnarray}
	is incompatible with the factorisability condition given by Eq.~(\ref{fact}) (shown in Appx.\ref{grhardy}). Similarly, the simultaneous validity of the CLL relations given by
	\begin{eqnarray} 
		\mathcal{P}_{CLL} &= P(+1,+1|\mathcal{A}_{1},\mathcal{B}_{1}, \kappa)- P(+1,+1|\mathcal{A}_{2},\mathcal{B}_{2}, \kappa) > 0 \label{c} \label{c1}\\[2.5pt]
		&P(-1,+1|\mathcal{A}_{2},\mathcal{B}_{1}, \kappa) = 0 \label{c2}\\[2.5pt]
		&P(+1,-1|\mathcal{A}_{1},\mathcal{B}_{2}, \kappa) = 0 \label{c3}
	\end{eqnarray}
	is also found to be incompatible with the factorisability condition Eq.~(\ref{fact}) (shown in Appx. \ref{grhardy} and \ref{grCLL}). Hence, the measurement outcome statistics satisfying the Hardy or CLL relations would signify the untenability of the assumption of predictability based on which Eq.~(\ref{fact}) is obtained, thereby providing an empirically validated certification of DI certified randomness. Thus, the logical basis for DI certification of randomness by invoking the Hardy or CLL relations is similar to that justifying the use of the CHSH inequality for the same purpose.
	
	Next, before proceeding to discuss the specifics of the quantitative evaluations of the bounds of the Hardy- and CLL-certified DI certified randomness, we briefly recall in the following section the relevant basics of this quantification issue.
	

	\section{Quantification of DI certified randomness in terms of min.-Entropy}\label{qcgr}
	
	In Information Theory, the quantity min-Entropy characterises the minimum unpredictability involved in the probability distribution \cite{renner}. In our treatment, we consider min-Entropy as the quantifier of certified randomness to facilitate a meaningful comparison of our results with those of the earlier relevant works where min-Entropy is considered as a quantifier of randomness.
	
	For a given amount of nonlocality, the amount of certified randomness ($\mathcal{R}_{xy}$) corresponding to $x^{th}$ choice of Alice's measurement and $y^{th}$ choice of Bob's measurement, optimising over all observed behaviour $\mathscr{P}_{obs}\equiv \{P(a, b | \mathcal{A}_{x}, \mathcal{B}_{y}, \kappa)\}$, is given by
	\begin{equation} \label{Rfd}
		\mathcal{R}_{xy} = - \log_2 \Big[\max\limits_{\{a,b, \mathscr{P}_{obs}\}} \ P(a, b | \mathcal{A}_{x}, \mathcal{B}_{y}, \kappa) \Big]  
	\end{equation}
	
	Here we note that the quantity $\mathcal{R}_{xy}$ defined in Eq.~(\ref{Rfd}) has explicit dependence on the measurement choices $x$ and $y$. Thus, there are two possible ways for evaluating the amount of DI certified randomness corresponding to a given amount of nonlocality - (i) one can minimise $\mathcal{R}_{xy}$ over all $x,y$ and obtain the minimum amount of DI certified randomness which we call the guaranteed amount ($\mathcal{R}_{g}$), and (ii) one can maximise $\mathcal{R}_{xy}$ over all $x,y$ and obtain the maximum achievable amount of DI certified randomness ($\mathcal{R}_{max}$). In the next section, first, we proceed for evaluating the guaranteed amount of randomness, i.e., $\mathcal{R}_{g}$, corresponding to Hardy and CLL relations.

	
	\section{Results: The Guaranteed amount of DI certified randomness based on the Hardy/Cabello-Liang-Li relations} \label{digr}
	
	Here we evaluate the quantity $\mathcal{R}_{g}$ which has the following precise operational meaning: For an arbitrarily prepared system and any combination of the pairs of measurement settings, if the statistics of joint measurement outcomes violate the Bell inequality or satisfy the 2-outcome Hardy/CLL relations, at least $\mathcal{R}_{g}$ bits amount of DI certified randomness is ensured for a given amount of nonlocality as signified by the non-zero values of Hardy or CLL parameter. This bound has particular importance in the context of cryptographic applications for ensuring the security of a random string under any adversarial guessing, irrespective of whether an adversary has access to information regarding the settings of the measurements performed by the user \cite{pironio2,bier}, and/or has control over the preparation procedure. It is using such a measure that one can guarantee a RNG to satisfy Shannon's version of Kerckhoffs's principle \cite{shan} which is a central tenet of modern cryptography viz. the requirement that a cryptographic system should be designed assuming that ``the enemy knows the system".
	
	Therefore, for a given amount of nonlocality, the minimum value of $\mathcal{R}_{xy}$ defined in Eq.~(\ref{Rfd}) corresponds to the guaranteed bound of DI certified randomness, $\mathcal{R}_{g}$, given as follows
	\begin{eqnarray} \label{rmindi}
		\mathcal{R}_{g}  &=& \min\limits_{\mathcal{A}_{x}, \mathcal{B}_{y}} \Big( \ \mathcal{R}_{xy} \ \Big) \nonumber \\
		&=& - \log_2 \Big[\ \max\limits_{\mathcal{A}_{x}, \mathcal{B}_{y}} \ 
		\max\limits_{\{a,b, \mathscr{P}_{obs}\}} \ P(a, b | \mathcal{A}_{x}, \mathcal{B}_{y}, \kappa)  \Big]  \nonumber \\
		&& \text{such that} \nonumber \\
		&&  \mathcal{P}_{Hardy}>0 \ [\text{or} \ \mathcal{P}_{CLL}>0 ]
	\end{eqnarray}
	
	A significance of the above expression given by Eq.~(\ref{rmindi}) lies in determining the upper bound on the probability of guessing the most probable pair of outcomes, which is given by $\frac{1}{2^{\mathcal{R}_{g}}}$, a quantity of key importance from the point of view of adversarial guessing \cite{renner, scabook, smith, cachin}.
	
	Now, in order to evaluate $\mathcal{R}_{g}$ by considering all possible observed behaviours, $\mathscr{P}_{obs}$, we first need to make an assumption of the theory which governs the realisation of such observed behaviours. To this end, we consider the following cases: (A) No-signalling theory \cite{Masanes2006} in which the set of behaviours necessarily obey the no-signalling conditions (as given by the Eqs.~(\ref{ns1}) and (\ref{ns2})), denoted as $\mathcal{NS}$. (B) Quantum theory, denoted as $\mathcal{Q}$.

	\subsection{$\mathcal{R}_{g}$ in NS theory} \label{digrns}
	
	Given that the procedure for certifying DI certified randomness discussed in Sec.~(\ref{certi}) hinges only on the no-signalling condition, it is natural to evaluate the guaranteed amount of certified randomness (corresponding to a given amount of nonlocality) against any adversarial guessing attack which is constrained only by the no-signalling principle.
	
	For this purpose, we consider the observed behaviour, $\mathscr{P}_{obs} \in \mathcal{NS}$. An important point to be noted is that a no-signalling set forms a polytope constituting a finite set of nonlocal and local deterministic vertices~\cite{barret, pivo}. In the 2-2-2 scenario, the $\mathcal{NS}$ polytope is eight dimensional and it has eight nonlocal and sixteen local deterministic vertices. In terms of bit variables $\tilde{x}=x-1,~\tilde{y}=y-1,~\tilde{a}=(1-a)/2,~\tilde{b}=(1-b)/2$, all these vertices can be represented succinctly as follows:
	\begin{eqnarray}
		\mbox{PR}^{\alpha\beta\gamma}(\tilde{a},\tilde{b}\vert \tilde{x},\tilde{y})\!\!&=&\!\!\frac{1}{2}\delta (\tilde{a}\oplus\tilde{b}=\tilde{x}\tilde{y}\oplus \alpha \tilde{x}\oplus \beta\tilde{y}\oplus \gamma), \\[2pt]
		\mbox{LD}^{\alpha_1\alpha_2\beta_1\beta_2}(\tilde{a},\tilde{b}\vert \tilde{x},\tilde{y})\!\!&=&\!\!\delta(\tilde{a},\alpha_1\tilde{x}\oplus\alpha_2)~\delta(\tilde{b},\beta_1\tilde{y}\oplus\beta_2),
	\end{eqnarray}
	where $\delta$ is Dirac's delta function, $\oplus$ denotes bit addition, $\alpha,\beta,\gamma \in\{0,1\}$, and $\alpha_1,\alpha_2,\beta_1,\beta_2\in \{0,1\}$.
	By exploiting the symmetries under (local) reversible relabelling of measurements and outcomes, it is sufficient to consider only those nonlocal no-signalling behaviours that can be expressed as a convex combination of one Popescu-Rohrlich box (PR Box) violating a CHSH inequality maximally, and 8 local deterministic (LD) distributions which saturate the local bound of the CHSH inequality \cite{bier16}. We consider the CHSH inequality in its standard form given by
	\begin{eqnarray}\label{chsh}
		\!\!\!\!\mathbf{B} \ &\equiv& \langle \mathcal{A}_{1} \mathcal{B}_{1}\rangle +\langle \mathcal{A}_{1} \mathcal{B}_{2}\rangle +\langle \mathcal{A}_{2} \mathcal{B}_{1}\rangle -\langle \mathcal{A}_{2} \mathcal{B}_{2}\rangle \leq 2,\\[2pt]
		&~& \mbox{where}~~ \langle \mathcal{A}_{x} \mathcal{B}_{y}\rangle =\sum_{a,b}  \ ab \ P(a,b\vert \mathcal{A}_{x}\mathcal{B}_{y}).\nonumber
	\end{eqnarray}
	Therefore, without loss of generality, we consider nonlocal no-signalling behaviours $\mathscr{P}_{obs} \in \mathcal{NS}$ which are expressed as a convex mixture of the one nonlocal vertex $\mathscr{P}_{PR}=\mbox{PR}^{000}(\tilde{a},\tilde{b}\vert \tilde{x},\tilde{y})$ satisfying $\mathbf{B}(\mathscr{P}_{PR})=4$, and eight local deterministic vertices  $\mathscr{P}_{LD_i}$ such that $\mathbf{B}(\mathscr{P}_{LD_i})=2$. Then,
	\begin{eqnarray}\label{pns}
		&~& \mathscr{P}_{obs} = q_0 \ \mathscr{P}_{PR} + \sum_{i=1}^{8} \ q_i \ \mathscr{P}_{LD_i}.
	\end{eqnarray}
	Here $q_0>0$, $q_i \geq 0 \ \forall  \ i= \{1,..,8\}$ and $q_0+\sum_{i=1}^8 q_i=1$.
	Note that the form of Hardy and CLL correlations we consider are also of the form $\mathscr{P}_{obs}$ given by Eq.~(\ref{pns}). It then follows that the corresponding Hardy and CLL nonlocality parameters are given by $\mathcal{P}_{Hardy}=\mathcal{P}_{CLL}=\frac{q_0}{2}$.\\\\
	
	Now, the maximum probability $\mathcal{P}^{\ast}(\mathscr{P}_{obs})$ is given as follows
	\begin{equation}
		\mathcal{P}^{\ast}(\mathscr{P}_{obs}) = \frac{q_0}{2} + (1-q_0) = 1-\frac{q_0}{2}
	\end{equation}
	Which, in turn, gives the NS-bounds of guaranteed randomness for both the cases of Hardy and CLL relations as follows
	\begin{eqnarray}
		(\mathcal{R}_{g})_{\mathcal{NS}}^{Hardy} &=& -log_2 \ \Big( 1-\mathcal{P}_{Hardy}\Big) \label{bns1}\\
		(\mathcal{R}_{g})_{\mathcal{NS}}^{CLL} &=& -log_2 \ \Big( 1-\mathcal{P}_{CLL}\Big) \label{bns2}
	\end{eqnarray}

	\subsection{$\mathcal{R}_{g}$ in quantum theory} \label{resq}
	Here we consider the observed behaviour, $\mathscr{P}_{obs} \equiv\{P(a,b|\mathcal{A}_{x},\mathcal{B}_{y},\rho_{AB})\} \in \mathcal{Q}$ and the joint probabilities $P(a,b|\mathcal{A}_{x},\mathcal{B}_{y},\rho_{AB}) = \Tr[\rho_{AB} M_{a|x} \otimes N_{b|y}]$, where $M_{a|x}$ and $N_{b|y}$ are elements of POVMs $\mathcal{A}_{x}\equiv\{M_{a|x}\}$ and $\mathcal{B}_{y}\equiv\{N_{b|y}\}$ respectively, satisfying $M_{a|x},N_{b|y} \geq 0$ and $\sum\limits_{x}M_{a|x}=1$, $\sum\limits_{y}N_{b|y}=1$. Now, without loss of generality, by applying Naimark’s dilation theorem \cite{Paulsen_2003}, one can consider the measurements $M_{a|x}$ and $N_{b|y}$ as projectors. Thus, from now on we write $M_{a|x}=\Pi_{a|x}$ and $N_{b|y}=\Pi_{b|y}$ with $(\Pi_{a|x})^2=\Pi_{a|x}$ and $(\Pi_{b|y})^2=\Pi_{b|y}$. Note that while the evaluation of the DI bound of $\mathcal{R}_{g}$ should be independent of the dimension of states and corresponding measurement operators, in the 2-2-2 scenario, by applying Jordan's lemma, we can always find a basis such that density matrix corresponding to the state is in block diagonal form with maxim block size $2\times2$ and the measurement operators has a decomposition such that each part acts only on $2\times2$ block of the density matrix. This in turn reduces the problem of dimension-independent evaluation of $\mathcal{R}_{g}$ to evaluating the optimal value of $\mathcal{R}_{xy}$ over all possible pure two-qubit states.
	
	Let us first consider the following general bipartite pure state shared between Alice and Bob:
	\begin{equation} \label{obsh}
		\ket{\psi}=c_{00} \ket{00}+c_{01}\ket{01}+c_{10}\ket{10}+c_{11}\ket{11}
	\end{equation}
	where $c_{ij}\geq 0$ and $\sum\limits_{i,j\in\{0,1\}} |c_{ij}|^2=1$. $\ket{0}$ and $\ket{1}$ are the eigenstate of the observable $\sigma_z=\ket{0}\bra{0}-\ket{1}\bra{1}$ with eigenvalues $+1$ and $-1$ respectively. The observables for Alice and Bob are given as follows
	\begin{eqnarray} \label{meas}
		&& \mathcal{A}_{1}=\ket{0}\bra{0}-\ket{1}\bra{1} \ ; \ \ \ \mathcal{A}_{2}=\ket{u_0}\bra{u_0}-\ket{u_1}\bra{u_1} \  ; \nonumber  \\
		&&\mathcal{B}_{1}=\ket{0}\bra{0}-\ket{1}\bra{1} \ ; \ \ \ \mathcal{B}_{2}=\ket{v_0}\bra{v_0}-\ket{v_1}\bra{v_1} \ ;
	\end{eqnarray}
	where $\ket{u_0}=\cos\frac{\alpha}{2}\ket{0}+e^{i \phi}\sin\frac{\alpha}{2}\ket{1}$; $\ket{u_1}=-\sin\frac{\alpha}{2}\ket{0}+e^{i \phi}\cos\frac{\alpha}{2}\ket{1}$; $\ket{v_0}=\cos\frac{\beta}{2}\ket{0}+e^{i \epsilon}\sin\frac{\beta}{2}\ket{1}$; $\ket{v_1}=-\sin\frac{\beta}{2}\ket{0}+e^{i \epsilon}\cos\frac{\beta}{2}\ket{1}$; $0\leq\alpha,\beta\leq \pi$ and $0\leq\phi,\epsilon\leq2\pi$. It is important to note here that without loss of generality, we fix observable $\mathcal{A}_{1}=\mathcal{B}_{1}=\sigma_z$ and we keep the other two observables $\mathcal{A}_{2}$ and $\mathcal{B}_{2}$ as well as the state $\ket{\psi}$ in most general form \cite{Rai2022}.

	\subsubsection{$\mathcal{R}_{g}$ in quantum theory by using the Hardy relations} \label{resqh}
	
	It has been shown \cite{Rai2022} that in order to satisfy the constraints on joint probabilities given by Eq.~(\ref{h1}-\ref{h4}), the state $\ket{\psi}$ and observables must satisfy $\langle\psi|u_1\otimes0\rangle=\langle\psi|0 \otimes v_1 \rangle=\langle\psi|u_0 \otimes v_0\rangle=0$. Then the three independent state parameters appearing in Eq.~(\ref{obsh}) are expressed in terms of the two measurement parameters $\alpha$ and $\beta$. Thus, the pure non-maximally entangled two-qubit states exhibiting Hardy nonlocality are of the following form
	\begin{equation}\label{hardystate}
		\ket{\psi}=\frac{1}{\sqrt{1+\tan^2\frac{\alpha}{2}+\tan^2\frac{\beta}{2}}}\qty(\tan\frac{\alpha}{2}\ket{u_0v_1}+\tan\frac{\beta}{2}\ket{u_1v_0}+\ket{u_1v_1})
	\end{equation}

	Thus, the joint probability distributions exhibiting Hardy-nonlocality are expressed as functions of two variables, say $s_1$ and $s_2$, with $s_1=\sin^2\frac{\alpha}{2}$, $s_2=\sin ^2\frac{\beta}{2}$, satisfying $0\leq s_1,s_2\leq 1$. Note that the quantum maximum value of $(\mathcal{P}_{Hardy})_{max}=\frac{5\sqrt{5}-11}{2}$ occurs when $s_1=s_2=\frac{\sqrt{5}-1}{2}$.
	
	Now, for a given amount of Hardy-nonlocality, in order to evaluate the guaranteed randomness $(\mathcal{R}_{g})$, we need to find the maximum joint probability. For this purpose, we write the joint probability distributions, denoted by $\mathscr{P}_{\tiny{H}}$, in terms of $s_1$ and $s_2$ as follows:
	\begin{eqnarray}
		\mathscr{P}_{\tiny{H}} \equiv 	
		\begin{array}{|c||c|c|c|c|}
			\hline 
			&(+,+)& (+,-)&(-,+)&(-,-) \\
			\hline \hline
			\mathcal{A}_{1}\mathcal{B}_{1} & \frac{(1-s_1)(1-s_2) s_1s_2}{1-s_1s_2} & \frac{(1-s_1) s_1s_2^2}{1-s_1s_2} & \frac{(1-s_2)s_1^2s_2}{1-s_1s_2} & 1-s_1s_2\\
			\hline 
			\mathcal{A}_{1}\mathcal{B}_{2} &\frac{(1-s_1)s_1s_2}{1-s_1s_2} & 0 & \frac{(1-s_1)^2s_2}{1-s_1s_2} &  \frac{1-s_2}{1-s_1s_2} \\
			\hline 
			\mathcal{A}_{2}\mathcal{B}_{1} & \frac{(1-s_2)s_1s_2}{1-s_1s_2} & \frac{(1-s_2)^2s_1}{1-s_1s_2} & 0 & \frac{1-s_1}{1-s_1s_2} \\
			\hline
			\mathcal{A}_{2}\mathcal{B}_{2} & 0 & \frac{(1-s_2)s_1}{1-s_1s_2} & \frac{(1-s_1)s_2}{1-s_1s_2} & \frac{(1-s_1)(1-s_2)}{1-s_1s_2} \\
			\hline 
		\end{array}
		\label{behahs}
	\end{eqnarray}
	with $\mathcal{P}_{Hardy}=\frac{(1-s_1)(1-s_2) s_1s_2}{1-s_1s_2}$. Note that it has been shown \cite{Rai2022} that the behaviour $\mathscr{P}_H$ given by Eq.~(\ref{behahs}) provides self-testing of pure non-maximally entangled states of the form given by Eq.~(\ref{hardystate}), along with the corresponding measurement settings for all $(s_1,s_2)\in(0,1)\times (0,1)$. 
	
	Since any distribution which leads to self-testing is an extremal point of the set of quantum correlations, any given value of Hardy's nonlocal parameter in the quantum range $P_{Hardy}\in (0, \frac{5\sqrt{5}-11}{2}]$ corresponds to a set of extremal distributions, all of which are useful for generating DI certified randomness. Thus, one obtains a range for the DI certified randomness even if Hardy's nonlocal parameter $P_{Hardy}$ has a specific value. Note that in order to generate such Hardy-certified DI randomness, the nonlocal parameter is not required to be maximum, unlike the Bell inequality based DI randomness.

\begin{figure}[ht]
		\centering
		\includegraphics[width=0.95\linewidth]{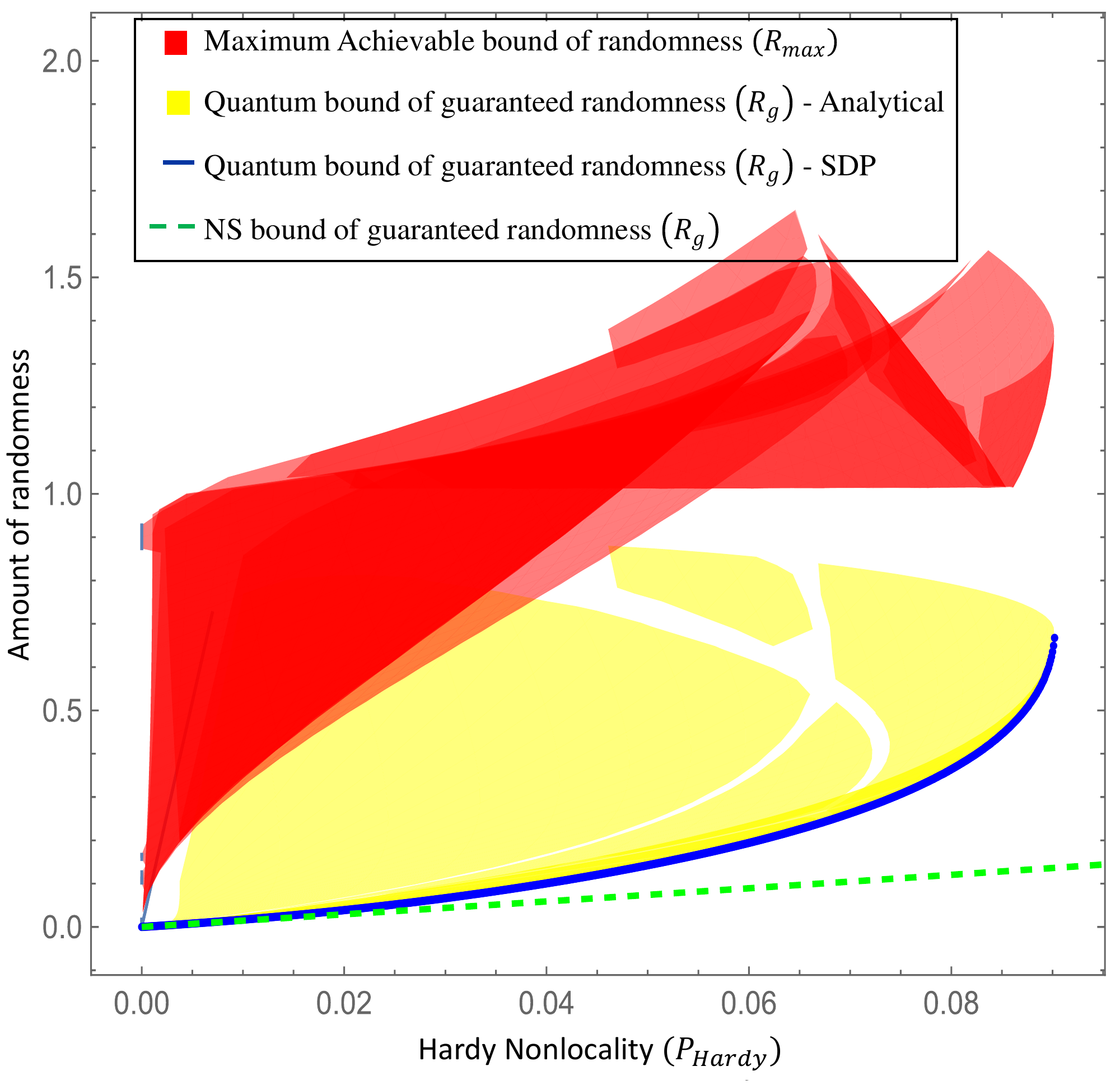}
		\caption{The yellow region represents the variations of guaranteed bounds of DI certified randomness as functions of Hardy-nonlocality signified by the Hardy parameter $\mathcal{P}_{Hardy}$ evaluated analytically in quantum theory. The blue ad green dashed-lines represent the SDP computed quantum bound of guaranteed randomness and analytically obtained NS bound of guaranteed randomness respectively. The red region represents the maximum achievable bound of randomness as a function of Hardy-nonlocality. Note that for each non-zero values of Hardy parameter ($P_{Hardy}$), there exists a set of extremal quantum behaviours and each such $P_{Hardy}$ corresponds to a set of $\mathcal{R}_g$ and $\mathcal{R}_{max}$. Thus, analytically obtained guaranteed and maximum bounds are represented by regions over varying amounts of nonlocality as quantified by $P_{Hardy}$. On the other hand, the SDP computed bound is found to be the lower bound of the guaranteed randomness.}
		\label{fighardyrmin}
	\end{figure}

	It is straightforward to show that the maximum joint probability, $\mathcal{P}^{\ast}_{H}$, corresponding to the behaviour given by Eq.~(\ref{behahs}) is given by, 
	\begin{equation}
		\mathcal{P}^{\ast}_{H} = \max \ \qty[(1-s_1s_2), \frac{1-s_2}{1-s_1s_2}, \frac{1-s_1}{1-s_1s_2}] \ \ \ \forall s_1,s_2 \in \{0,1\}
	\end{equation}
	Therefore, the quantum mechanically evaluated Hardy-certified $\mathcal{R}_{g}$ is given by $\mathcal{R}_{g}^{Hardy}=-\log_{2}[\mathcal{P}^{\ast}_{H}]$. We have illustrated (Fig.~(\ref{fighardyrmin})) the variation of analytically obtained values of $\mathcal{R}_{g}^{Hardy}$ corresponding to different values of Hardy-nonlocality parameter. Further, this evaluation is done by employing the SDP technique \cite{Pironio2010, SDP1, SDP2}. We, then, compare such SDP computed bound with the analytically obtained bound. It is observed that the SDP computed bound provides the lower bound of analytically obtained $\mathcal{R}_{g}^{Hardy}$ (see Fig.~(\ref{fighardyrmin})).
	
	\subsubsection{$\mathcal{R}_{g}$ in quantum theory by using the CLL relations} \label{resqCLL}
	
	Similar to the Hardy relations, here we characterise the joint probability distributions exhibiting CLL nonlocality. The constraints on joint probabilities given by Eq.~(\ref{c1}-\ref{c3}) imply that $\langle\psi|u_1\otimes0\rangle=\langle\psi|0 \otimes v_1\rangle=0$. This, in turn, specifies two of the independent state parameters appearing in Eq.~(\ref{obsh}) in terms of the two measurement parameters $\alpha$ and $\beta$. Then, the form of pure non-maximally entangled two-qubit states exhibiting CLL nonlocality has the following form (up to multiplication by some global phase) \cite{Rai2021}
		\begin{eqnarray}\label{statehfCLL}
			\ket{\psi}&=& e^{i\delta}\sqrt{1-c\qty(1+\tan^2\frac{\alpha}{2}+\tan^2\frac{\beta}{2})} \ \ket{u_0v_0} \nonumber \\
			&& + \sqrt{c} \ \qty( \tan\frac{\alpha}{2} \ \ket{u_0v_1}+ \tan\frac{\beta}{2} \ \ket{u_1v_0} + \ket{u_1v_1} )
		\end{eqnarray}
		where $0\leq c\leq \frac{1}{1+\tan^2\frac{\alpha}{2}+\tan^2\frac{\beta}{2}}$. The maximum value of $\mathcal{P}_{CLL}$ given by $(\mathcal{P}_{CLL})^{opt}_{Q}=0.1078$ occurs for the particular state with $c=0.3068$ and $\delta=\pi$, when Alice and Bob measure in the same direction given by $\alpha=\beta=1.6136$ radian.
		
		Now, in order to evaluate the analytical bound of guaranteed randomness $(\mathcal{R}_{g}^{CLL})$, we need to find the maximum joint probability. The behaviour exhibiting CLL nonlocality is given as follows:
		\begin{eqnarray} 
			\nonumber
			\mathscr{P}_{\tiny {CLL}} \equiv
			\begin{array}{|c||c|c|c|c|}
				\hline 
				&(+,+)& (+,-)&(-,+)&(-,-) \\
				\hline \hline
				\mathcal{A}_{1}\mathcal{B}_{1} & p & y p & x p & 1- x p \\
				&&&& -yp-p \\
				\hline 
				\mathcal{A}_{1}\mathcal{B}_{2} &p(1+y) & 0 & 1-c(1+x)& c(1+x) \\
				&&&-p(1+y)& \\
				\hline 
				\mathcal{A}_{2}\mathcal{B}_{1} & p(1+x) & 1-c(1+y) & 0 & c(1+y) \\
				&&-p(1+x)&&\\
				\hline
				\mathcal{A}_{2}\mathcal{B}_{2} & 1-cx & c x & c y  & c  \\
				&-cy-c&&& \\
				\hline 
			\end{array}
			\label{behahmaxCLL}\\
		\end{eqnarray}
		where $p=\frac{1-2 c \cos\delta \sqrt{xy \{\frac{1}{c}-(1+x+y)\}}+c (1+x+y-xy)}{(x+1) (y+1)}$, $x=\tan^2\frac{\alpha}{2}$, $y=\tan^2\frac{\beta}{2}$. The CLL parameter is given by $\mathcal{P}_{CLL}=p-1+c(1+x+y)$. 
		
		Now, in order to find the maximum joint probability, $\mathcal{P}^{\ast}_{CLL}$, corresponding to the considered behaviour given by Eq.~(\ref{behahmaxCLL}), we proceed as follows. Let $\mathcal{P}^{\ast}(\mathcal{A}_i,\mathcal{B}_j)=\max\limits_{a,b}P(a,b|\mathcal{A}_i,\mathcal{B}_j)$ be the maximum joint probability corresponding to the choice of each pair of measurement settings $(i,j)$. Now, for all $P_{CLL}>0$, due to symmetry, it can be seen that $\mathcal{P}^{\ast}(\mathcal{A}_1,\mathcal{B}_1)=\mathcal{P}^{\ast}(\mathcal{A}_2,\mathcal{B}_2)$ and $\mathcal{P}^{\ast}(\mathcal{A}_1,\mathcal{B}_2)=\mathcal{P}^{\ast}(\mathcal{A}_2,\mathcal{B}_1)$. Moreover, it is straightforward to see that $\mathcal{P}^{\ast}(\mathcal{A}_1,\mathcal{B}_2)>\mathcal{P}^{\ast}(\mathcal{A}_1,\mathcal{B}_1)$. Therefore, the maximum probability corresponding to the behaviour given by Eq.~(\ref{behahmaxCLL}) is given by
		\begin{equation} \label{pmax1CLL}
			\mathcal{P}^{\ast}_{CLL}=\max\limits_{x,y,c,\delta} \Big[p(1+y), 1-c(1+x)-p(1+y),c(1+x)\Big] 
		\end{equation}
		
\begin{figure}[ht]
		\centering
		\includegraphics[width=0.9\linewidth]{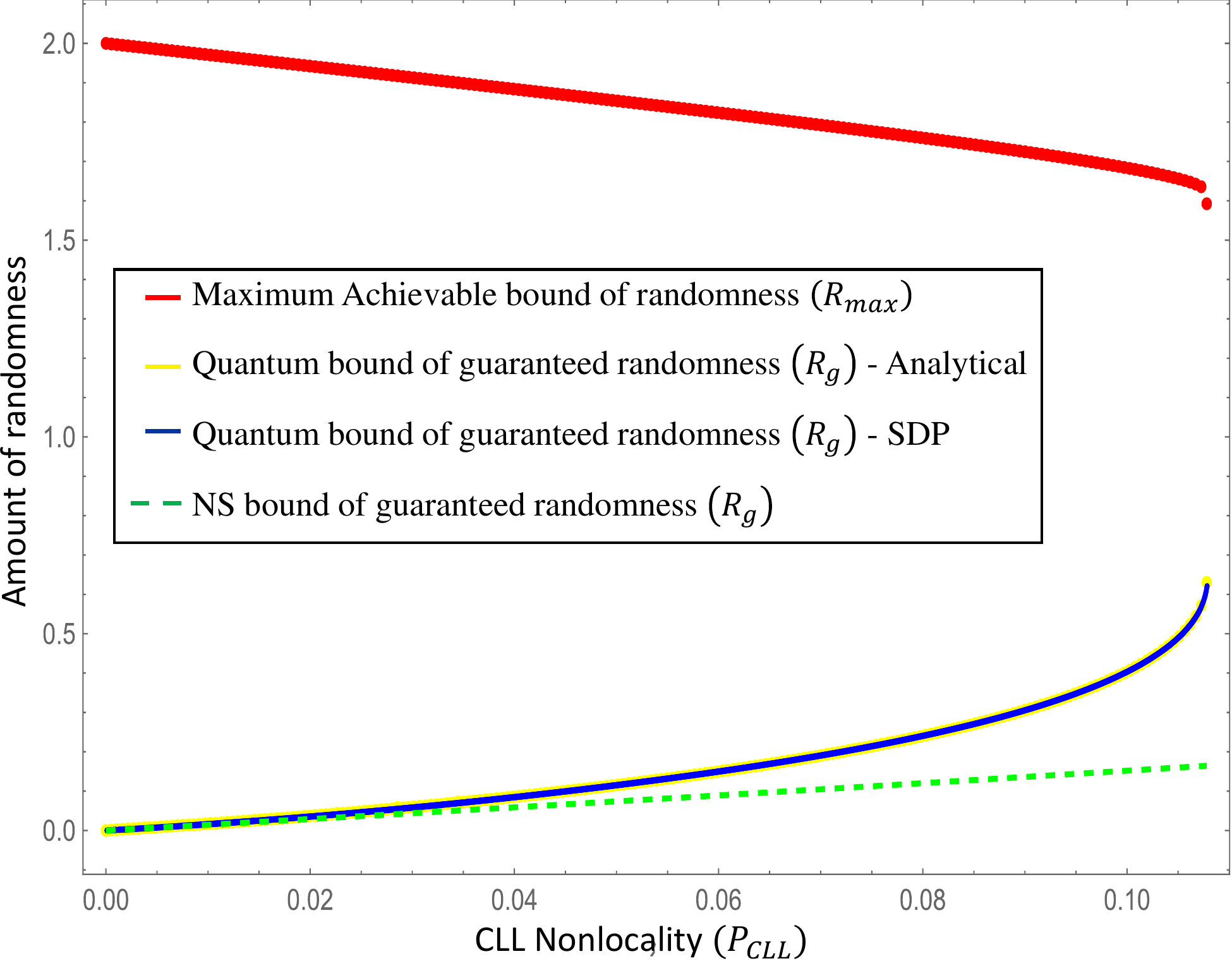}
		\caption{ These curves represent the variation of CLL-certified randomness with CLL-nonlocality. The yellow curve represents the variations of guaranteed bounds of DI certified randomness as functions of CLL-nonlocality signified by the CLL parameter $\mathcal{P}_{CLL}$ evaluated analytically in quantum theory. Blue and green dashed-lines represent the SDP computed quantum bound of guaranteed randomness and analytically obtained NS bound of guaranteed randomness respectively. Note that the SDP computed quantum bound is found to be the same as that obtained analytically. The red curve represents the maximum achievable bound of randomness as a function of CLL-nonlocality. It is seen that close to 2 bits of certified randomness can be achieved for small amount of nonlocality, thereby demonstrating the incommensurability between randomness and nonlocality.}
		\label{figCLLrmin}
	\end{figure}

		Thus, the quantum mechanically evaluated bound of CLL-certified $\mathcal{R}_g$ is given by $\mathcal{R}^{CLL}_g = -\log_2\qty[\mathcal{P}^{\ast}_{CLL}]$. The variation of such guaranteed bound with CLL nonlocality has been illustrated in Fig.~(\ref{figCLLrmin}), along with the bound of guaranteed randomness computed using the SDP technique. Here we note that, unlike the Hardy case, the guaranteed bound $\mathcal{R}^{CLL}_g$ is found to be unique for each non-zero value of $\mathcal{P}_{CLL}$. 
	
	\section{Results: Maximum amount of DI certified randomness based on the Hardy/Cabello-Liang-Li relations} \label{reshardymax}
	
	Here we evaluate the maximum achievable bounds of DI certified randomness that can be certified by using the Hardy and CLL relations. For a given amount of nonlocality, the maximum value of $\mathcal{R}_{xy}$ defined in Eq.~(\ref{Rfd}) corresponds to the maximum achievable bound of DI certified randomness, $\mathcal{R}_{max}$, is given as follows
	\begin{eqnarray} \label{rmaxf}
		\displaystyle \mathcal{R}_{max}  &=& \max\limits_{\mathcal{A}_{x}, \mathcal{B}_{y}} \Big( \ \mathcal{R}_{xy} \ \Big) \nonumber \\
		&=& - \log_2 \Big[\min\limits_{\{\mathcal{A}_{x},\mathcal{B}_{y}\}} \ \max\limits_{\{a,b, \rho\}} \ P (a, b | \mathcal{A}_x, \mathcal{B}_y, \rho)\Big] \nonumber \\
		&& such \ that \nonumber \\
		&&  \mathcal{P}_{Hardy}>0 \ [or \ \mathcal{P}_{CLL}>0 ]
	\end{eqnarray}
	
	Now, for any behaviour, the maximum randomness corresponds to the distributions of equally likely events. In the two party-two output scenario, there are four possible events corresponding to each pair of measurement settings. Thus, maximum of 2 bits of randomness can be possible to certify in such scenario. 
	
	\subsection{Maximum amount of DI certified randomness based on the Hardy relations}
	
	Hardy relations imply that for each of these three measurement pairs $\qty(\mathcal{A}_1,\mathcal{B}_2)$, $\qty(\mathcal{A}_2,\mathcal{B}_1)$ and $\qty(\mathcal{A}_2,\mathcal{B}_2)$, the occurrence of one particular event (i.e., a pair of outcomes) is ruled out (Eqs.~(\ref{h2}-\ref{h4})). Thus, corresponding to each of these three pairs of measurement settings, maximal randomness occurs when the remaining three events occur with equal probability $\frac{1}{3}$. 
	
	First, considering the pair of measurement settings $\qty(\mathcal{A}_1,\mathcal{B}_2)$ or $\qty(\mathcal{A}_2,\mathcal{B}_1)$, it is straightforward to obtain that the joint probabilities will be $\frac{1}{3}$ when $s_1=\frac{1}{2}$ and $s_2=\frac{4}{5}$. For such values of $s_1$ and $s_2$, the value of the Hardy parameter is $P_{Hardy}(s_1=\frac{1}{2}, s_2=\frac{4}{5})=\frac{1}{15}\approx 0.0667$ which is less than the maximum value of $\frac{5\sqrt{5}-11}{2}\approx 0.0902$. Thus, the maximum amount of randomness that can be certified for the pair of measurement settings $\qty(\mathcal{A}_1,\mathcal{B}_2)$ or $\qty(\mathcal{A}_2,\mathcal{B}_1)$ is $-\log_2\qty[\frac{1}{3}]\approx 1.5850$ bits corresponding to non-maximal violation of Hardy nonlocality.
	
	Next, for the pair of measurement settings $\qty(\mathcal{A}_2,\mathcal{B}_2)$, the joint probabilities will be $\frac{1}{3}$ when $s_1=s_2=\frac{1}{2}$. For such values of $s_1$ and $s_2$, the value of he Hardy parameter is $P_{Hardy}(s_1=s_2=\frac{1}{2})=\frac{1}{12}\approx 0.0833$ which is again less than the maximum value.
	
	Now, interestingly, for the remaining pair of measurement settings $\qty(\mathcal{A}_1,\mathcal{B}_1)$, since $0<P(+1,+1|\mathcal{A}_1,\mathcal{B}_1)\leq\frac{5\sqrt{5}-11}{2}$, there is a possibility that other three events can occur with equal probability less than $\frac{1}{3}$. These events $(+1,-1)$, $(-1,+1)$ and $(-1,-1)$ will occur with same probability if $P(+1,-1|\mathcal{A}_1,\mathcal{B}_1)=P(-1,+1|\mathcal{A}_1,\mathcal{B}_1)=P(-1,-1|\mathcal{A}_1,\mathcal{B}_1)$. This then fixes the values of the parameters $s_1=s_2=0.8295$. The corresponding Hardy parameter is $P_{Hardy}(s_1=s_2=0.8295)=0.0641$ and the amount of certified randomness is $-\log_2\qty[0.3119]\approx 1.6807$ bits. Note that this is the maximum achievable amount of randomness that can be device-independently certified using the Hardy relations (see Fig.~(\ref{fighardyrmin})).   
	
	\subsection{Maximum amount of DI certified randomness based on the CLL relations}
	
CLL relations has three constraints on joint probability distributions, out of which two constraints are $P(+1,-1|\mathcal{A}_1,\mathcal{B}_2)=0$ and $P(-1,+1|\mathcal{A}_2,\mathcal{B}_1)=0$. The other constraint is the CLL nonlocality parameter, which unlike the Hardy nonlocality parameter, is given by the difference between two joint probabilities, $0<P(+1,+1|\mathcal{A}_1,\mathcal{B}_1)-P(+1,+1|\mathcal{A}_2,\mathcal{B}_2)<0.1078$. This, in turn, gives rise to the possibility of having $P(a,b|\mathcal{A}_i,\mathcal{B}_{j=1}) \to \frac{1}{4} \ \forall a,b,i=j$. Let us first analyse the case when $P(a,b|\mathcal{A}_1,\mathcal{B}_1) \to \frac{1}{4}$. In order to have such probability distribution corresponding to the choice of measurements $(\mathcal{A}_1,\mathcal{B}_1)$, $p=\frac{1}{4}$ and $x=y=1$. This, in turn, fixes the value of the parameter $c=\frac{\cos^2\delta}{3+\cos^2\delta}$, if $\cos\delta\geq0 \ \implies 0\leq \delta\leq\frac{\pi}{2}$. Moreover, $0<\mathcal{P}_{CLL}\leq 0.1078 \ \implies \ \frac{1}{4}<c\leq 0.2859$. Now in this domain, the only solution that satisfies both $\mathcal{P}_{CLL}>0$ and $p=\frac{1}{4}$ when $c\to\frac{1}{4}$. Thus, close to 2 bits of randomness can be certified with $\mathcal{P}_{CLL}\to 0$.
		
		Next, we analyse the case when $P(a,b|\mathcal{A}_2,\mathcal{B}_2) \to \frac{1}{4}$. Here, $c=\frac{1}{4}$ and $x=y=1$. This, in turn, fixes the value of the parameter $p=\frac{1}{8}(3-\cos\delta)$, if $0\leq \delta\leq2\pi$. Moreover, $0<\mathcal{P}_{CLL}\leq 0.1078 \ \implies \ \frac{1}{4}<p\leq 0.3578$. Now in this domain, $\delta\in(0,1.4328)\cup(4.8504,6.2832)$ (all the angles are in radian). Note that in such value of $c=\frac{1}{4}$, the joint probability $P(-1,+1|\mathcal{A}_1,\mathcal{B}_2)=P(+1,-1|\mathcal{A}_2,\mathcal{B}_1)=\frac{1}{2}-2 p$, thus in turn fixes the range of $0\leq p\leq \frac{1}{4}$. Therefore, in this case also close to 2 bits of randomness can be certified with $\mathcal{P}_{CLL}\to 0$ when $p\to\frac{1}{4}$ and $c\to\frac{1}{4}$. Thus it is indeed possible to achieve close to 2 bits of certified randomness with arbitrary small amount of CLL certified nonlocality. 
		
		Now, from the behaviour given by Eq.~(\ref{behahmaxCLL}), it follows that $\mathcal{P}^{\ast}(\mathcal{A}_i,\mathcal{B}_(j\neq i))>\mathcal{P}^{\ast}(\mathcal{A}_i,\mathcal{B}_{j=1})$. Thus, the maximum amount of randomness given as follows:
		\begin{eqnarray}
			\mathcal{R}_{max}=-\log_2\qty[\max\limits_{x,y,c,\delta} \Big[1-c(1+x+y), cx,cy,c\Big] ] \label{rmaxCLLf}
		\end{eqnarray}
		The variation of the maximum amount with different non-zero values of $P_{CLL}$ has been illustrated in Fig.~(\ref{figCLLrmin}).


	\section{Concluding Remarks} \label{con}
	
	As the basis for this work, we have first provided justification of the way the Hardy and Cabello-Liang-Li relations can enable the DI certification of randomness, based on only the no-signalling condition at the statistical level. 
	
Then, focusing on the quantitative evaluation of DI certified randomness, for a given amount of Hardy nonlocality, we find that the amount of DI certified randomness is not unique. This implies a range of DI certified randomness for a given amount of Hardy nonlocality. The underlying reason is that, for a given non-zero value of the Hardy nonlocality parameter, all the intersection points between the hyperplanes formed by the Hardy relations and the quantum mechanical joint-probability space form a set of quantum extremal distributions, giving rise to a range of DI certified randomness. It is to be noted here that for the Hardy relations, the extremality of such distributions is ensured by the self-testing argument \cite{Rai2022}. In contrast, for the CLL relations, we find that the amount of DI certified randomness is unique, similar to the earlier results for the CHSH nonlocality. The variations of the obtained DI certified randomness with the amounts of Hardy and CLL-nonlocality are illustrated in Figs.~(\ref{fighardyrmin}) and (\ref{figCLLrmin}) respectively. Note that the occurrence of a range of DI certified randomness for a given amount of Hardy nonlocality is in sharp contrast to the results of the DI randomness studies based on CHSH or tilted-Bell inequalities. Therein only the maximum violations of the relevant inequalities have been proved to be quantum extremal, and, thus, a given amount of CHSH value corresponds to a unique amount of randomness certified by Bell inequalities. Furthermore, for both the Hardy and CLL relations, the matching of the lower bounds of the analytically obtained guaranteed amount with the respective SDP computed bounds signify the tightness of our analytical treatment. 
	
	Now, considering the maximum achievable bound of DI certified randomness, it is analytically shown that both for small amounts of Hardy- and CLL-nonlocality, larger amounts of randomness are realised (see red coloured regions in Figs.~ \ref{fighardyrmin} and \ref{figCLLrmin}). A particularly significant result is that in the simplest 2-2-2 scenario, it is possible to realise close to the maximum amount of 2 bits of CLL-certified randomness for a range of pure non-maximally entangled states, even for small amounts of CLL-nonlocality. Therefore, this demonstrates the incommensurability between the maximum achievable bound of randomness, nonlocality and entanglement in a single setup. 
	
It will be interesting to extend this line of study by using other forms of local realist inequalities, such as different forms of the higher settings Bell inequality \cite{ms1, ms2, ms3, ms4, brunner2014, chain1, chain2, chain3, GHZ, deng, avis}, or the generalised variants of the Hardy relations \cite{msh1,msh2}. Another possible direction of study could be to go beyond the 2-2-2 scenario using the recently suggested measure of nonlocality which has been invoked to argue for ensuring the quantitative compatibility between entanglement and nonlocality for arbitrary dimensional system \cite{mn1, mn2}. 
	
	To conclude, the upshot of the results of this work is the reinforcement of the realisation of a fundamental feature of the quantum world which is linked with randomness. While the certification of DI certified randomness necessarily requires nonlocality, the nature of the quantitative relationship between them is more nuanced than what has been discussed earlier.


	\section*{Acknowledgements}
	SS acknowledges the financial support from INSPIRE programme, Department of Science and Technology, Govt. of India. SG acknowledges the KVPY scholarship funded by the Department of Science and Technology, Govt. of India. AR acknowledges support from the National Research Foundation of Korea (Grant No. NRF-2021R1A2C2006309, NRF-2022M1A3C2069728) and the Institute for Information \& Communication Technology Promotion (IITP) (the ITRC Program/IITP-2023-2018-0-01402). DH thanks NASI for the support provided by Senior Scientist Platinum Jubilee Fellowship and acknowledges support of the QuEST-DST Project Q-98 of the Govt. of India. US acknowledges the support provided by Ministry of Electronics $\&$ Information Technology (MeitY), Government of India under grant for ``Centre for Excellence in Quantum Technologies" with Ref. No. 4(7)/2020-ITEA.  
	
	\appendix
	
	\section{Certification of DI certified randomness using the 2-outcome Hardy relations} \label{grhardy}
	
	Let us consider the two-outcome Hardy relations characterised by the simultaneous validity of the following four conditions on joint probabilities 
	\begin{eqnarray} 
		P(+1,+1|\mathcal{A}_{1},\mathcal{B}_{1}, \kappa) &=& \mathcal{P}_{Hardy} > 0 \label{ha1}\\[2.5pt]
		P(-1,+1|\mathcal{A}_{2},\mathcal{B}_{1}, \kappa) &=& 0 \label{ha2}\\[2.5pt]
		P(+1,-1|\mathcal{A}_{1},\mathcal{B}_{2}, \kappa) &=& 0 \label{ha3}\\[2.5pt]
		P(+1,+1|\mathcal{A}_{2},\mathcal{B}_{2}, \kappa) &=& 0 \label{ha4}
	\end{eqnarray}
	
	Now, applying the factorisability condition given in the text by Eq.~(\ref{fact}) to the above mentioned Hardy relations, we obtain
	\begin{eqnarray} 
		P(+1|\mathcal{A}_{1}, \kappa ) ~P(+1|\mathcal{B}_{1}, \kappa) &=& \mathcal{P}_{Hardy}  > 0 \label{ha5} \\[2.5pt]
		P(-1|\mathcal{A}_{2}, \kappa ) ~P(+1|\mathcal{B}_{1}, \kappa) &=& 0    \label{ha6}\\[2.5pt]
		P(+1|\mathcal{A}_{1}, \kappa ) ~ P(-1|\mathcal{B}_{2}, \kappa) &=& 0    \label{ha7}\\[2.5pt]
		P(+1|\mathcal{A}_{2}, \kappa ) ~ P(+1|\mathcal{B}_{2}, \kappa) &=& 0    \label{ha8}
	\end{eqnarray}
	
	Next, we show that the simultaneous validity of the above four Eqs.~(\ref{ha5})-(\ref{ha8}) is inconsistent with the factorisability condition. Specifically, we show the inconsistency of Eq.~(\ref{ha5}) with Eqs.~(\ref{ha6}-\ref{ha8}). For this purpose, we rewrite Eqs.~(\ref{ha6}) and (\ref{ha7}) respectively as follows
	\begin{eqnarray}
		P(+1|\mathcal{B}_{1}, \kappa) &=& P(+1|\mathcal{B}_{1}, \kappa)~P(+1|\mathcal{A}_{2}, \kappa ) \label{ha6new}\\[2.5pt]
		P(+1|\mathcal{A}_{1}, \kappa) &=& P(+1|\mathcal{A}_{1}, \kappa)~P(+1|\mathcal{B}_{2}, \kappa ) \label{ha7new}
	\end{eqnarray}
	Multiplying the above two equations leads to the following
	\begin{eqnarray}
		&&P(+1|\mathcal{A}_{1}, \kappa)~ P(+1|\mathcal{B}_{1}, \kappa) \nonumber \\[2.5pt]
		=&& P(+1|\mathcal{A}_{1}, \kappa) P(+1|\mathcal{B}_{1}, \kappa) P(+1|\mathcal{A}_{2}, \kappa) P(+1|\mathcal{B}_{2}, \kappa) \label{ha6plus7}
	\end{eqnarray}
	Finally, using Eq.(\ref{ha8}) in Eq.(\ref{ha6plus7}), we obtain
	\begin{equation}
		P(+1|\mathcal{A}_{1}, \kappa)~ P(+1|\mathcal{B}_{1}, \kappa)=0 
	\end{equation}
	which contradicts Eq.~(\ref{ha5}), i.e., the condition that $\mathcal{P}_{Hardy}>0$. Hence, the simultaneous validity of all the conditions imposed on the four joint probabilities given by Eqs.~(\ref{ha1} - \ref{ha4}) is inconsistent with the factorisability condition given by Eq.~(\ref{fact}) in the text. This implies violation of the condition of predictability. Thus, the Hardy relations can be employed for certifying DI certified randomness.
	
	\section{Certification of DI certified randomness using the Cabello-Liang-Li relations} \label{grCLL}
	
	Here we consider a variant of the Hardy relations, namely, the CLL relations which have also been used for showing  \cite{cabello, liang} quantum nonlocality independent of the Bell type inequalities. In the following, we will show that the simultaneous validity of all the CLL relations contradicts the factorisability condition given by Eq.~(\ref{fact}).
	
	The CLL relations can be written as follows
	\begin{eqnarray} 
		\mathcal{P}_{CLL} &= P(+1,+1|\mathcal{A}_{1},\mathcal{B}_{1}, \kappa)- P(+1,+1|\mathcal{A}_{2},\mathcal{B}_{2}, \kappa) > 0 \label{ca1}\\[2.5pt]
		&P(-1,+1|\mathcal{A}_{2},\mathcal{B}_{1}, \kappa) = 0 \label{ca2}\\[2.5pt]
		&P(+1,-1|\mathcal{A}_{1},\mathcal{B}_{2}, \kappa) = 0 \label{ca3}
	\end{eqnarray}
	
	Now, applying the factorisability condition given by Eq.~(\ref{fact}), Eqs.~(\ref{ca2}) and (\ref{ca3}) can be rewritten respectively as
	\begin{eqnarray}
		P(+1|\mathcal{B}_{1}, \kappa) &=& P(+1|\mathcal{B}_{1}, \kappa)~P(+1|\mathcal{A}_{2}, \kappa ) \label{ca2new}\\[2.5pt]
		P(+1|\mathcal{A}_{1}, \kappa) &=& P(+1|\mathcal{A}_{1}, \kappa)~P(+1|\mathcal{B}_{2}, \kappa ) \label{ca3new}
	\end{eqnarray}
	
	It then follows from the above two Eqs.~(\ref{ca2new}) and (\ref{ca3new})
	\begin{eqnarray}
		&~&P(+1|\mathcal{A}_{1}, \kappa)~ P(+1|\mathcal{B}_{1}, \kappa)- P(+1|\mathcal{A}_{2}, \kappa)~ P(+1|\mathcal{B}_{2}, \kappa) \nonumber \\[2.5pt]
		&=& P(+1|\mathcal{A}_{2}, \kappa)~ P(+1|\mathcal{B}_{2}, \kappa)~\left\{-1+P(+1|\mathcal{A}_{1}, \kappa)~ P(+1|\mathcal{B}_{1}, \kappa)\right\} \nonumber\\[2.5pt]
		&\leq& 0 \nonumber \\ [2.5 pt]
		&\Rightarrow& \mathcal{P}_{CLL} \leq 0\label{ca23new}.
	\end{eqnarray}
	thereby contradicting Eq.(\ref{ca1}). Hence, any joint probability distribution of the measurement outcomes satisfying all the CLL relations given by Eqs.~(\ref{ca1}-\ref{ca3}) would be inconsistent with the factorisability condition (Eq.~(\ref{fact}) in the text). This implies violation of the condition of predictability. Thus, the CLL relations can be employed for certifying DI certified randomness, similar to the use of the Hardy relations.

	\section{Computation of guaranteed bounds of DI certified randomness in quantum theory} \label{A}
	
	For computing the guaranteed bound of DI certified randomness for both quantum mechanically and using the no signalling (NS) principle, we proceed as follows.
	
	Let us consider that $S$ is any convex subset of the set of joint conditional probability distributions $\vec{\mathscr{P}}=\Big\{P(a,b|x,y): a,b\in \{\pm 1\}~\mbox{and}~x,y \in\{1,2\}\Big\}$. We further assume that all the elements in $S$ satisfy the NS condition. Then the guaranteed amount of randomness, say $R^{S}_{g}$, that can be certified, subject to a given nonlocality condition, is given by
	\begin{eqnarray} \label{sdpr}
		R^{S}_{g} &=&\min\limits_{\vec{\mathscr{P}}~\in~S}\left[-\mbox{log}_{2}\left(\max\limits_{a,b,x,y}~P(a,b|x,y)\right)\right] \nonumber \\
		&=&-\mbox{log}_{2}\left[\max\limits_{\vec{\mathscr{P}}~\in~S}\left(\max\limits_{a,b,x,y}~P(a,b|x,y)\right)\right] \nonumber \\
		&&subject~to \nonumber \\
		&&relevant~constraints~on \ P(a,b|x,y)
	\end{eqnarray}
	
	This optimisation problem can readily be solved by applying the semi-definite-programming (SDP) technique as this is a case of the convex optimisation problem. 
	
	Note that the phrase ``relevant constraints on $P(a,b|x,y)$" used in Eq.~(\ref{sdpr}) is explained as follows: First, in order to evaluate the NS bound of $\mathcal{R}_{g}$, we consider the subset $S \in \{P(a,b|x,y)\}$ satisfying either the Hardy or CLL relations. Secondly, for obtaining the quantum mechanically computed lower bound of $\mathcal{R}_{g}$, we apply the specific quantum theoretic-constraints\footnote{(i) $P (a, b | x, y, \rho) = Tr \ [ M_{a|\mathcal{A}_{x}} \otimes M_{b|\mathcal{B}_{y}} \ \rho ] $, (ii) $\rho$ $\in$ $H_A \otimes H_B$ of dimension $d_A d_B$ and (iii) each of the measurements $\mathcal{A}_{x}$ of Alice corresponds to a positive-operator-valued-measure (POVM): $\mathcal{A}_{x}=\{M_{a|\mathcal{A}_{x}}\}_a$ with $M_{a|\mathcal{A}_{x}} \geq 0$ for all $a$ and $\sum_a M_{a|\mathcal{A}_{x}}= \mathbb{I}_{d_A}$. Similarly, each measurement setting $\mathcal{B}_{y}$ of Bob corresponds to a positive-operator-valued-measure (POVM): $\mathcal{B}_{y}=\{M_{b|\mathcal{B}_{y}}\}_b$ with $M_{b|\mathcal{B}_{y}} \geq 0$ for all $b$ and $\sum_b M_{b|\mathcal{B}_{y}}= \mathbb{I}_{d_B}$.   } on the subset $S$.
	
	Now, for solving the optimisation problem, given by Eq.~(\ref{sdpr}), using the SDP technique \cite{SDP1, SDP2}, we choose our convex set $S$ as different levels of the NPA-Hierarchy, denoted here by $Q^{(k)}$ where $k\in\{0,1,\mbox{1+ab}, 2, 3,...\}$. All these different levels are convex, and form a sequence of outer approximations of the set of quantum behaviours $Q$, i.e., $Q^{(0)}\supseteq Q^{(1)}\supseteq ... Q^{(k)}...\supseteq Q$. Note that the zeroth level approximation $Q^{(0)}$ is the set of all NS behaviours. Also, note that in the 2-2-2 scenario, the convergence of $Q^{(k)}$ is very fast so that at the level 1+ab (an intermediate level which lies between the levels 1 and 2), $Q^{(1+ab)}\simeq Q$. Thus, in order to compute the guaranteed \textit{quantum} bound of DI certified randomness from the Hardy relations, the following SDP sub-problem is solved:
	\begin{subequations}\label{main1}
		\begin{align}
			&\max P(a,b|x,y);\\
			&~subject~to\nonumber \\
			&~\vec{\mathscr{P}} \in Q^{(1+ab)}, \\
			&~P(+1,+1~|~1,1)= \mathcal{P}_{Hardy}, \\
			&~P(-1,+1~|~2,1)=0, \\
			&~P(+1,-1~|~1,2)=0,  \\
			&~P(+1,+1~|~2,2)=0. 
		\end{align}
	\end{subequations}
	
	Next, for computing the guaranteed \textit{NS} bound of certified DI certified randomness from the Hardy relations, the following SDP sub-problem is solved:
	\begin{subequations}\label{main12}
		\begin{align}
			&\max P(a,b|x,y);\\
			&~\mbox{subject~to}\nonumber \\
			&~\vec{\mathscr{P}} \in Q^{0}, \\
			&~P(+1,+1~|~1,1)= \mathcal{P}_{Hardy}, \\
			&~P(-1,+1~|~2,1)=0, \\
			&~P(+1,-1~|~1,2)=0,  \\
			&~P(+1,+1~|~2,2)=0. 
		\end{align}
	\end{subequations}
	
	Similarly, in order to compute the guaranteed bounds on DI certified randomness from the CLL relations, the following SDP sub-problems are solved: \\
	
	\textit{quantum} guaranteed bound of DI certified randomness:
	\begin{subequations}\label{main2}
		\begin{align}
			&~\max P(a,b|x,y); \\
			&~\mbox{subject~to}\nonumber \\
			&~\vec{\mathscr{P}} \in Q^{(1+ab)}, \\
			&~P(+1,+1~|~1,1)-P(+1,+1~|~2,2)= \mathcal{P}_{CLL}, \\
			&~P(-1,+1~|~2,1)=0, \\
			&~P(+1,-1~|~1,2)=0. 
		\end{align}
	\end{subequations}
	
	\textit{NS} guaranteed bound of DI certified randomness:
	\begin{subequations}\label{main22}
		\begin{align}
			&~\max P(a,b|x,y); \\
			&~\mbox{subject~to}\nonumber \\
			&~\vec{\mathscr{P}} \in Q^{0}, \\
			&~P(+1,+1~|~1,1)-P(+1,+1~|~2,2)= \mathcal{P}_{CLL}, \\
			&~P(-1,+1~|~2,1)=0, \\
			&~P(+1,-1~|~1,2)=0. 
		\end{align}
	\end{subequations}

	After solving these sub-problems, the computation of $\mathcal{R}_{g}$ readily follows: For any given fixed value of $\mathcal{P}_{Hardy}$ ($\mathcal{P}_{CLL}$) we find the maximum values of $P(a,b|x,y)$ for all 16 possible choices from $a,b\in \{\pm 1\}$ and $x,y\in\{1,2\}$. We then select the maximum from the resulting 16 values, denoted by $\mathcal{P}^{\ast}_{Hardy}$ ($\mathcal{P}^{\ast}_{CLL}$). Finally, the quantum mechanically computed guaranteed bounds of DI certified randomness are given by
	\begin{eqnarray}
		\mathcal{R}^{Q}_{g}(\mbox{Hardy}) &=&-\mbox{log}_{2}\left(\mathcal{P}^{\ast}_{Hardy}\right), \\
		\mathcal{R}^{Q}_{g}(\mbox{CLL}) &=&-\mbox{log}_{2}\left(\mathcal{P}^{\ast}_{CLL}\right).
	\end{eqnarray}
	
	Similarly, the NS guaranteed bounds of DI certified randomness have also been computed.

	\bibliography{references}

\end{document}